# Design and Implementation of Location and Activity Monitoring System Based on LoRa

**Shengwei Lin[1], Ziqiang Ying[1] and Kan Zheng[1]**
[1] Intelligent Computing and Communication Lab,
Key Lab of Universal Wireless Communications, Ministry of Education,
Beijing University of Posts and Telecommunications, Beijing, China, 100876.
[e-mail: shengweilin@bupt.edu.cn]

## *Abstract*

The location and human activity are usually used as one of the important parameters to monitor the health status in healthcare devices. However, nearly all existing location and monitoring systems have the limitation of short-range communication and high power consumption. In this paper, we propose a new mechanism to collect and transmit monitoring information based on LoRa technology. The monitoring device with sensors can collect the real-time activity and location information and transmit them to the cloud server through LoRa gateway. The user can check all his history and current information through the specific designed mobile applications. Experiment was carried out to verify the communication, power consumption and monitoring performance of the entire system. Experimental results demonstrate that this system can collect monitoring and activity information accurately and provide the long rang coverage with low power consumption.

***Keywords:*** LoRa, Activity monitoring, Location

## 1. Introduction

**W**ith the development of Internet of Things (IoT) communication technologies, more and more smart devices begin to be enjoyed by the people in our society [1] [18]. The environment and human status can be monitored through different kinds of methods. For example, the location information and activity can be collected by using location and movement sensors in the smart devices for healthcare applications [2]. Then, such information can be further explored to monitor the health status of human or animals.

At present, smart devices for monitoring the location and human or animal activity commonly adopt the short-range communication technologies such as Bluetooth, Wi-Fi to transmit the data to the cloud or mobile phone of the given user [3] [19]. However, these communication systems either have the limited coverage or have too high power consumption, which are not suitable for a few specific IoT applications, e.g., monitoring animal in large farms or ranches. Thus, the Low Power Wide Area Network (LPWAN) technologies become very promising to be used in smart devices [4]. There exists various LPWAN technologies and solutions, such as Sigfox, NB-IoT and LoRaWAN [13] [14] [15]. Among all of them, Sigfox and NB-IoT network are managed by local operator. Data transmission from sensors are avaliable only if subscription fee is paid [20]. Furthermore, there are limitation of data traffic and packet size [21].

LoRa is a wireless communication technology patented by Semtech [9]. Anyone who respects the protocol specifications can uses it free. The technology is presented in two parts - LoRa, the physical layer and LoRaWAN, the upper layers. LoRa physical layer adopts a special Chirp Spread Spectrum (CSS) modulation technology which significantly increases the communication range compared with Frequency Shift Keying (FSK) modulation. LoRaWAN is a MAC layer protocol and system architecture design. It adopts simple star network topology and easily to be deployed [10]. The combination of LoRa and LoRaWAN enables very-long-range transmissions with low power consumption. Consequently, some smart devices [11] [12], such as environment monitoring system [5] and smart irrigation system [6], have been implement based on LoRa technology.

In this paper, we design and implement location and activity monitoring system based on LoRa communication technology. The system mainly consists of cloud, LoRa gateway, LoRa device and application in mobile phone. The LoRa device can periodically collect the location and activity data, then send them to the LoRa gateway, which can forward data to the cloud. Meanwhile, the user can check the activity and location data through mobile applications.

This paper focuses on the design and implementation of the LoRa device for monitoring. We implement the hardware of LoRa device, including positioning module, step counter module, transmission module, power management module and control module. Moreover, we propose a mechanism to collect and transmit monitoring information with high accuracy and low power consumption. Finally, the experiments are carried out to demonstrate both the communication and monitoring performances.

The rest of this paper is presented as follows: Section 2 briefly introduces the overall architecture of the system. Section 3 describes the design and implementation of the proposed system including the hardware and software. In Section 4, we provide the experimental results of the proposed system. Finally, the conclusion is given in Section 5.

## 2. System Architecture

**Fig. 1** gives the architecture of the location and activity monitoring system based on LoRa, which mainly consists of four parts: 1) Cloud; 2) LoRa Gateway; 3) LoRa Device; 4) Mobile Application.

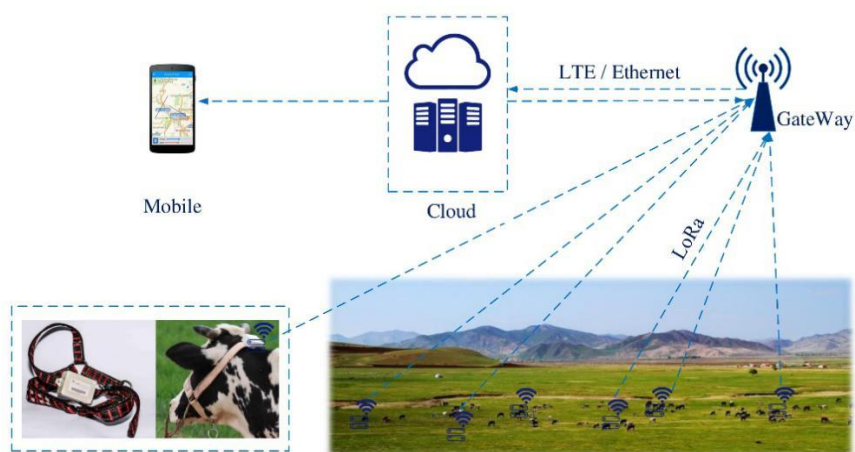

**Fig. 1**. Illustration of System Architecture

### 2.1 Cloud

The cloud is mainly used for data storage and analysis, and provide APIs for applications. There are a few function modules in the cloud, e.g., protocol transfer module, application service module and database. Protocol transfer is to extract application data from reported data encapsulated according to the LoRaWAN protocol [7] [22]. In order to provide various service, not only HTTP but MQTT servers are implemented in the application service module. Mobile application obtains data through the APIs of HTTP service module. Instant communication between device and cloud is implemented through MQTT protocol for message publishing and subscribing. Regarding the data storage, NoSQL databases such Redis are applied in the cloud because they have higher I/O speed than SQL database.

### 2.2 LoRa Gateway

The LoRa gateway is used for data forwarding between cloud and LoRa devices [6]. In the uplink, the gateway demodulates RF signal from device and forwards the data to the cloud through the UDP protocol. In the downlink, the gateway firstly parses the downlink UDP packets from the cloud, then transmit data to LoRa devices through LoRa communication links [16].

**2.3 LoRa Device**

The LoRa device is one of the key parts of this system. It collects the data of activity and location information, then transmits them to the LoRa gateway [17]. In this system, the device works in Class A mode, which is the lowest power consumption mode in LoRa systems. Moreover, the power consumption in the LoRa devices can be further reduced by adaptive data rate (ADR) mechanisms [16].

**2.4 Mobile**

Mobile application is responsible for the visualization of activity and location data. It can get analysis result of location and activity data through application programming interfaces (APIs) provided by cloud. Through mobile application, users can know the movement track and movement trend of animal or human.

## 3. System design and implementation

In this section, we only present the design and implementation of the LoRa device, which is the main contribution of this paper. Not only the hardware but also the software are designed and implemented by ourselves.

**3.1 Design on LoRa Device**

Location and activity information can be reflected by GPS and step information. In order to collect location information and the number of steps, we apply the positioning module and step counter module in LoRa device as shown in **Fig. 2**. Besides these modules, the LoRa transmission is taken care of by communication module. As usual, the control and power management modules exist in our LoRa device.

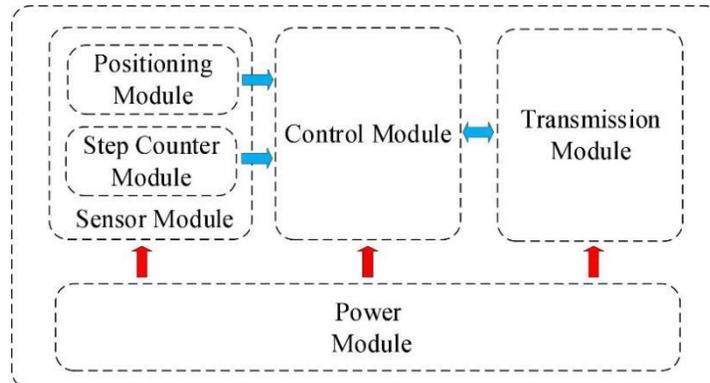

**Fig. 2**. Illustration of main modules in LoRa device

### 3.2 Implementation on LoRa Device

#### 3.2.1 Hardware Implementation

According to the design above, we have selected and implemented the hardware needed for all the modules. The details are given as below.

a) Positioning Module: The ublox's MAX-7Q module is used to get the location information. This module supports GPS and GLONASS, whose positioning error is within 10 meter. The location information can be correctly extracted in 30 second after the device is powered on.
b) Step Counter Module: ADXL345, a 3-axis acceleration sensor, is selected for monitoring the activity. The numbers of steps can be calculated by triaxial acceleration data. Control module can read the $x$, $y$, $z$ triaxial acceleration data acquired by ADXL345 via I$^2$C or SPI bus.
c) Communication Module: Semtech's SX1276 chip is chosen as LoRa modulation and demodulation module. The RF frequency of this chip ranges from 137 *MHz* to 1050 *MHz*. SX1276 has high receiver sensitivity for the long distance reliable transmission.
d) Control Module: The Microcontroller Unit (MCU) of STM32L151 is used to achieve the necessary control functions. The processor has rich peripheral interfaces, i.e., multiple groups of UART, SPI, I$^2$C. For the sake of low power consumption, STM32L151 can reduce the current when enters to standby mode.
e) Power Management Module: TPS61025 is adopted as the core chip. Its input voltage ranges from 0.9 *V* to 6.5 *V*. It provides stable 3.3 *V* voltage output for other modules by adjusting the corresponding voltage divider resistance. Moreover, solar panel is equiped with to charge lithium battery.

#### 3.2.2 Software Implementation

The software implementation on device mainly consists of collection and transmission on location and activity data. Besides, a mobile application is developed to better display location and movement track information. The details are given as below.

a) Collection on location and activity data
- Data Acquisition and Transmission

Data acquisition mainly includes step number acquisition and GPS information collection. ADXL345 is a three axis acceleration sensor with a 32 level FIFO cache which can be used to store three axis acceleration data. The MAX-7Q can send GPS raw information directly to the serial port. Considering the characteristic of these sensors, we have finished the program design of data acquisition and transmission. First, MCU wakes up and initializes. After that, MCU reads the 32 sets of acceleration data stored in FIFO and calculates the number of steps according to the step counting algorithm. Then, the step register is updated. Finally, MCU enters into standby mode. After a period of time, MCU continues to read the acceleration data. After one hour, MCU extracts GPS information to parse longitude and latitude, then packet it with 6 set of steps. Finally, the data is transmitted to the server through LoRa. The flowchart of this procedure is shown in **Fig. 3**.

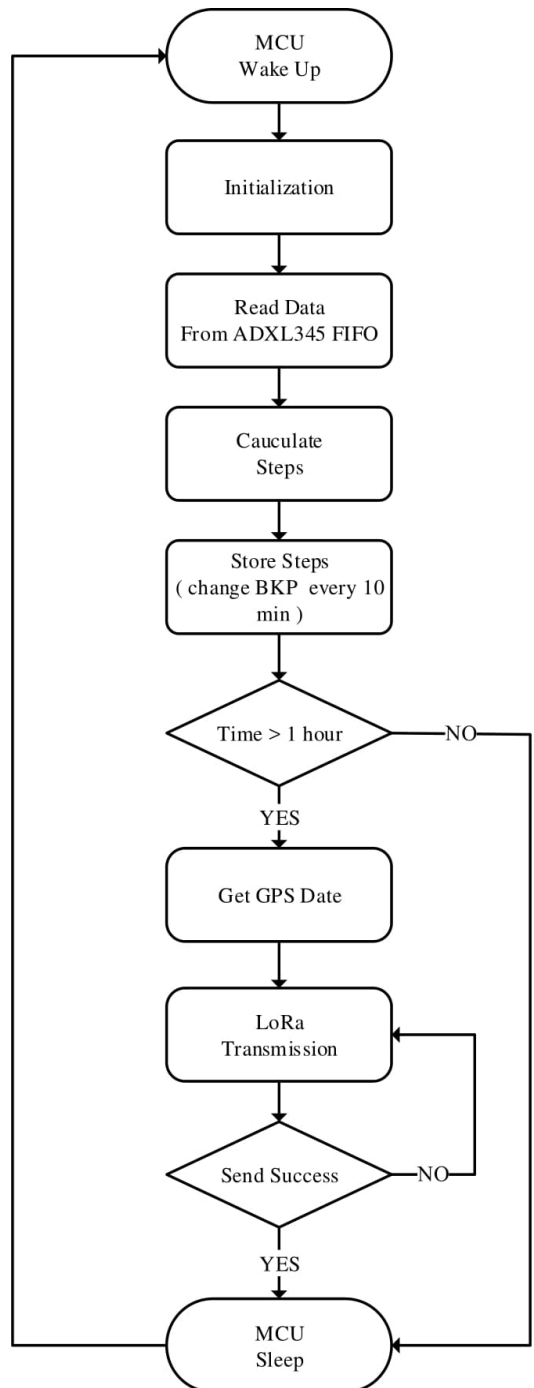

**Fig. 3**. Work flow of data acquisition and transmission

- Step counting algorithm

The step counting algorithm is based on the acceleration data of *x*, *y* and *z* collected by the ADXL345. The "Zero Crossing" method is used to calculate the step number [8]. When the device is placed correctly, the z axis of the ADXL345 sensor is perpendicular to the ground, and the x axis and the *y* axis are horizontal to the ground. During the movement of walking, the acceleration of *z* axis changes most. Therefore, the steps can be calculated through the periodic variation of *z* axis acceleration. The details of the implementation is explained as follows.

STEP 1. Dynamic threshold: Collect 32 sets of *z* axis acceleration data and calculate average $\alpha$, maximum $\beta$, minimum $\gamma$. The dynamic threshold *T* can be obtained by (1)

$$T = Max\{(\beta - \alpha), (\alpha - \gamma)\} \tag{1}$$

If *T* is less than a fixed value *S*, it may be judged to be static and step counter is set to 0, then the process is finished.

STEP 2. Precision setting: $p = T/m$ ($m$ is a fixed value).

STEP 3. Steps Count: The linear shift register contains 2 registers. *A* and *B* register respectively record previous and current acceleration data. If it's satisfied with (2), step counter might be increased by

$$\begin{cases} A - \alpha > p \\ B - \alpha < -p \end{cases} \tag{2}$$

STEP 4. Repeat STEP 3 until all data has been processed.

b) Mobile applications

In order to well display the activity and location information to users, we have developed an Android application for the mobile phone. Through this application, we can analyze the activity and location information collected by the device and show it to the user, the main functions are shown in **Fig. 4**, including: 1) the replay of motion trajectory; 2) the electronic fence; 3) high precision and accurate positioning; 4) the number of steps display at each time period. Through the replay work of motion trajectory, you can playback the historical movement path of the device according to the timeline. The replay function is also shown in **Fig. 4**.

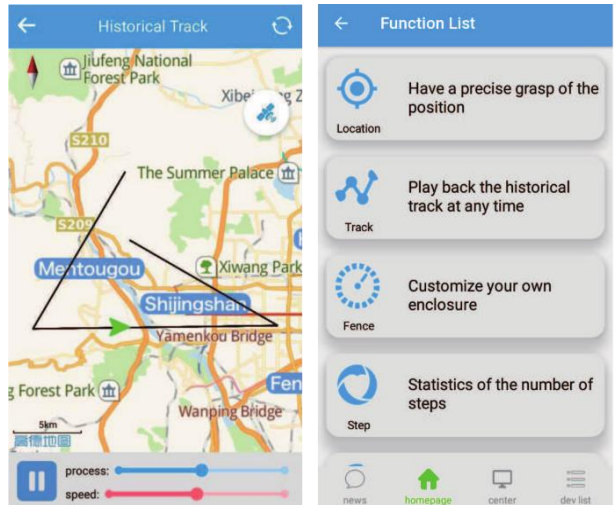
**Fig. 4**. GUI examples of mobile application

## 4. Experimental results and analysis

To verify the reliability of system, a plenty of experiments have been carried out to test the performances of LoRa communication, power consumption and monitoring.

### 4.1 Experimental configuration

The experiments for testing communication and monitoring performances happen in the suburb of Beijing. Due to the limitation of environment, experiment cannot be carried out 3 *km* farther away from gateway. The antenna is set up on an 8 meter high mast and the gateway works at 433 *MHz*. The experiment environment is relatively open, with some buildings below 3 floors. From the start point to 3 *km* away from the gateway, the device sends the number of steps and GPS position information to the gateway continuously. The experimental environment and test path are shown in **Fig. 5**.

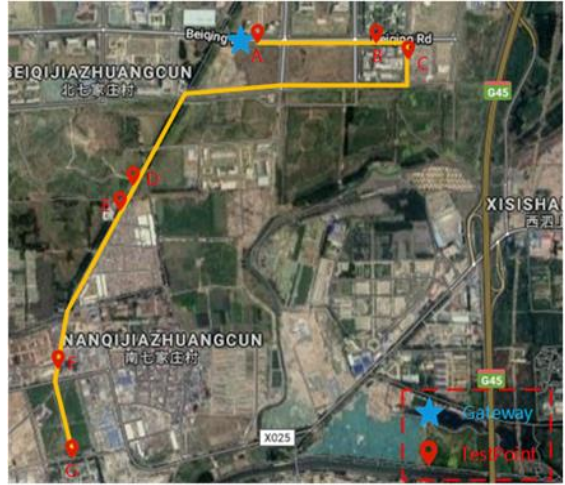

**Fig. 5**. Test environment and path

To test the power consumption performance, voltage regulator is prepared to record the working current in a single duty cycle and ammeter is prepared to record sleep current when device is in standby mode.

### 4.2 Experiment and analysis

### 4.2.1 LoRa Communication performance

In order to test the communication performance of LoRa under different conditions. On one hand, we measure the Received Signal Strength Indicator (RSSI) and Signal-Noise Ratio (SNR) with the power of 20 *dBm* and different Spreading Factor (SF) range from SF7 to SF12. The test results are shown in **Fig. 6** and **Fig. 7**. On the other hand, we test the effect of different transmit power on RSSI and SNR at the data rate of SF12. The test results are shown in **Fig. 8** and **Fig. 9**.

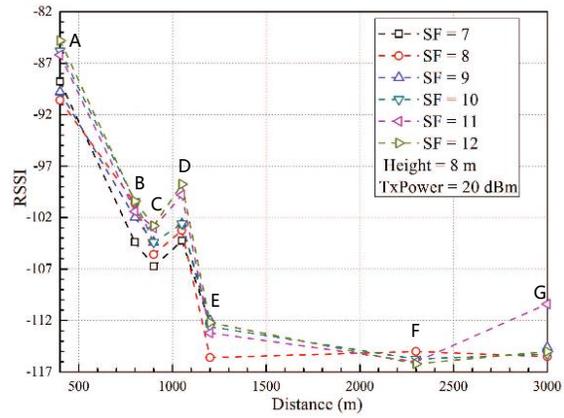

**Fig. 6**. RSSI with different SF

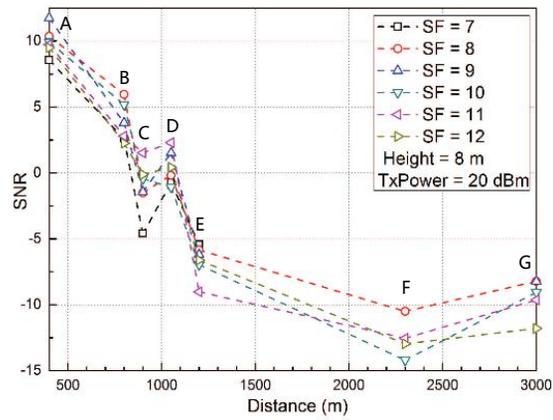

**Fig. 7**. SNR with different SF

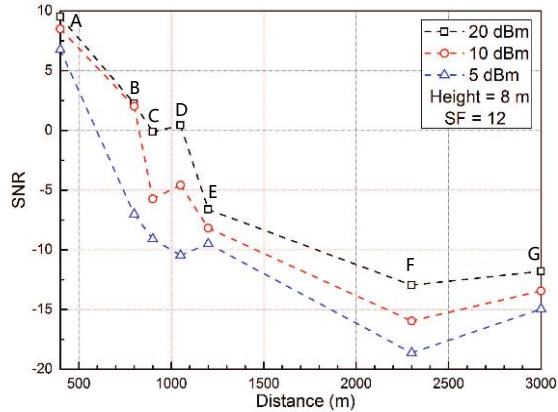

**Fig. 8**. SNR with different TxPower

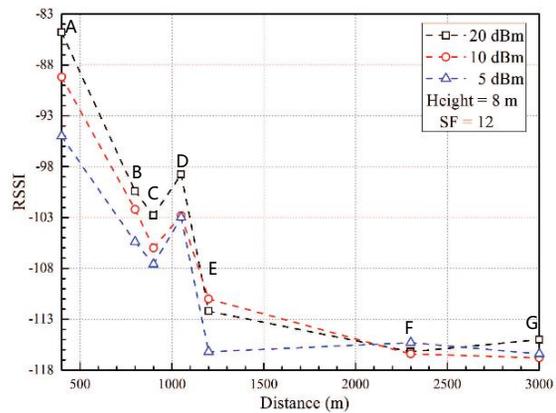

**Fig. 9**. RSSI with different TxPower

As a whole, the test results show that SF is positively correlated with RSSI and SNR, the transmitting power is also positively correlated with RSSI and SNR. With the communication distance increasing, RSSI keeps decreasing. However, it can be seen that RSSI and SNR on TestPoint D is better than TestPoint C. By analyzing **Fig. 5**, we found that there are some tall buildings around TestPoint C, while there are almost no obstructions between gateway and TestPoint D. The test result can be explained by line of sight communication. when the device is 3 *km* away from the gateway, it still has good communication performance. It is possible to provide the reliable transmission with larger distance.

### 4.2.2 Power Consumption Performance

To test the power consumption performance, device is configured to the following parameters:
- Class of MAC : Class A
- Bandwidth : 125 *kHz*
- Coding Rate : 4/5
- Payload Size : 16 bytes
- Transmitting Power : 20 *dBm*
- Spreading Factor : SF7 ~ SF12

Different SFs affect the duration of sending period, thus affect the power consumption of device. Through the voltage regulator and adding breakpoints in the program, Duration of sending period and working current is measured with different SFs as show in **Table 1**.

**Table 1**. Time On Air and Current with different SFs

| Spreading Factor | Time On Air [*ms*] | Current [*mA*] |
|---|---|---|
| SF12 | 1650 | 134 |
| SF11 | 910 | 134 |
| SF10 | 420 | 134 |
| SF9 | 230 | 134 |
| SF8 | 120 | 134 |
| SF7 | 70 | 134 |

Taking SF12 as example, we obtain the working current of the device in different period, as shown in **Fig. 10**.

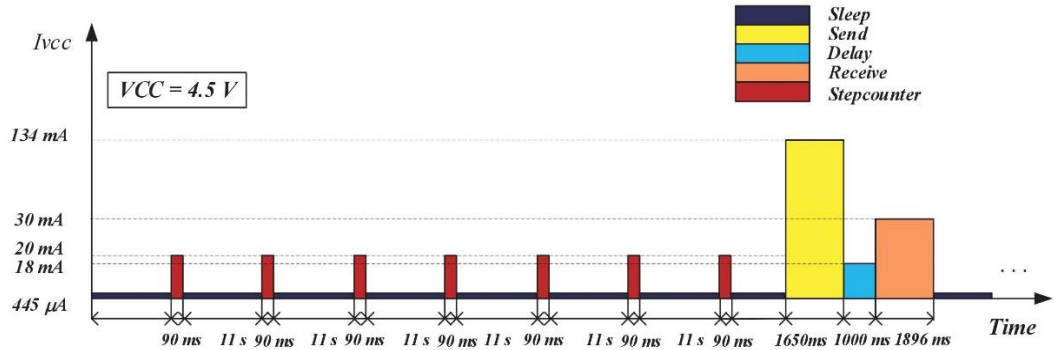

**Fig. 10**. Power consumption in different period

Therefore, the power consumption $Q$ in a single duty cycle can be calculated by

$$Q = m * (Qc + Qs) + Qt \qquad (3)$$

where *m* is number of times that device is in Stepcounter and Sleep period in a single duty cycle, $Qc$ is the power consumption of Stepcounter period, $Qs$ is the power consumption of Sleep period, $Qt$ is the sum power consumption of Send, Delay and Receive period. If LoRa device transmits data once an hour, the power consumption is about 18 *mAh* in one day. Solar panel can provide about 40 *mA* current in the day time. Thus, only charging half an hour every day, which provides 20 *mAh*, can keep the entire system working all days.

### 4.2.3 Monitoring Performance

The ADXL345 can be set in different sampling frequency. Normally, the human or animal's step frequency does not exceed 3 *Hz*. Thus, 6 *Hz* sampling frequency is sufficient to record the variation of z axis acceleration. we choose 3 *Hz* and 6 *Hz* sampling frequency to verify the accuracy of the number of steps. Our experiment method is to set 100, 200, 300 and 400 steps as the unit of measurement and calculate the average error at different frequencies. The experiment results are shown in **Table 2**.

Table 2. Average error at different frequency

| Steps \ Sampling frequency | 6 *Hz* | 3 *Hz* |
|---|---|---|
| 100 | 3 % | 11 % |
| 200 | 8 % | 14 % |
| 300 | 6 % | 17 % |
| 400 | 5 % | 10 % |

From the test results, it can be seen that the error rate is about 10 % - 20 % at 3 *Hz* sampling frequency, 3 % - 8 % for 6 *Hz*, respectively. Therefore, the number of steps can be collected with the acceptable error range if either 6 *Hz* or 3 *Hz* is set.

## 5. Conclusion

In this paper, the location and activity monitoring system based on LoRa communication has been designed and implemented. Through the device with step counter module and positioning module, the activity and location information can be collected with the satisfied accuracy. The data can be transmitted to cloud through LoRa communication. Moreover, users can playback the movement track and historical number of steps by mobile applications. Experiments on devices show that data can be transmitted reliably 3 *km* farther away from gateway and devices can work all day with solar panel. Therefore, this system can help us track and analyze the healthy information.

## Acknowledgment

This work was supported by the National Nature Science Foundation of China (Grant No. 61671089).